\documentclass[9pt]{article}
\usepackage[ngerman,english]{babel}
\usepackage[utf8]{inputenc}
\pdfoutput=1
\usepackage{amsmath}    
\usepackage{graphicx}   
\usepackage{verbatim}   
\usepackage{color}      
\usepackage{subfigure}  
\usepackage{hyperref}   
\usepackage{authblk}
\usepackage{mathtools}

\DeclarePairedDelimiter\ket{\lvert}{\rangle}
\DeclarePairedDelimiterX\braket[2]{\langle}{\rangle}{#1 \delimsize\vert #2}

\usepackage[numbers,compress]{natbib}

\usepackage{gensymb}
\usepackage[left=1.5cm,right=1.5cm,top=1.5cm,bottom=2cm]{geometry}

\begin{document}
\bibliographystyle{unsrtnat}
\renewcommand{\bibfont}{\small}

\title{Remote sensing of geomagnetic fields and atomic collisions in the mesosphere}
\author[1]{Felipe Pedreros Bustos \footnote{Corresponding author. Email: pedreros@uni-mainz.de}}
\author[2]{Domenico Bonaccini Calia}
\author[1]{Dmitry Budker}
\author[3]{Mauro Centrone}
\author[4]{Joschua Hellemeier}
\author[4]{Paul Hickson}
\author[2]{Ronald Holzlöhner}
\author[5]{Simon Rochester}

\affil[1]{Johannes Gutenberg University, Helmholtz Institut Mainz, Germany}
\affil[2]{European Southern Observatory, Germany}
\affil[3]{INAF – Osservatorio Astronomico di Roma, Italy}
\affil[4]{University of British Columbia, Canada}
\affil[5]{Rochester Scientific LLC, El Cerrito, CA}

\date{\today}

\maketitle

\begin{abstract}
Magnetic-field sensing has contributed to the formulation of the plate-tectonics theory, the discovery and mapping of underground structures on Earth, and the study of magnetism in other planets. Filling the gap between space-based and near-Earth observation, we demonstrate a novel method for remote measurement of the geomagnetic field at an altitude of 85--100 km. The method consists of optical pumping of atomic sodium in the upper mesosphere with an intensity-modulated laser beam, and simultaneous ground-based observation of the resultant magneto-optical resonance when driving the atomic-sodium spins at the Larmor precession frequency. The experiment was carried out at the Roque de Los Muchachos Observatory in La Palma (Canary Islands) where we validated this technique and remotely measured the Larmor precession frequency of sodium as 260.4(1) kHz, corresponding to a mesospheric magnetic field of 0.3720(1) G. We demonstrate a magnetometry accuracy level of 0.28 mG/$\sqrt{\text{Hz}}$ in good atmospheric conditions. In addition, these observations allow us to characterize various atomic-collision processes in the mesosphere. Remote detection of mesospheric magnetic fields has potential applications such as mapping of large-scale magnetic structures in the lithosphere and the study of electric-current fluctuations in the ionosphere.   
\end{abstract}


\section{Introduction}

Laser excitation of the atomic sodium layer, located between 85 and 100 km altitude in the upper mesosphere, allows astronomers to create artificial light sources, known as Laser Guide Stars (LGS), to assist adaptive optics systems \citep{Happer94}. A laser beam tuned to a wavelength resonant with the $3S_{1/2} \rightarrow 3P_{3/2}$ transition in sodium produces atomic fluorescence that is collected at ground with a telescope for real-time compensation of atmospheric turbulence in astronomical observations. Since the introduction of this technique \citep{Thompson87,Humphreys91}, research has been conducted to optimize laser excitation schemes in order to maximize the flux of photons returned to the ground. This technological progress has also catalyzed new concepts of laser remote sensing of magnetic fields with mesospheric sodium \citep{Higbie2011}. Because of the proximity of the sodium layer to the D and E regions of the ionosphere (between 70 km to 120 km altitude) mesospheric magnetometry opens the possibility to map local current structures in the dynamo region \citep{Yamazaki2016,Blanc80}. In addition, the capability of continuously monitoring the geomagnetic field at altitudes of 85--100 km could provide valuable information for modeling the geomagnetic field, detection of oceanic currents \citep{Tyler2003}, and for mapping and identification of large-scale magnetic structures in the upper mantle \citep{Maus2008}.

In a laser magnetometer, atoms are optically polarized, and the effects of the interaction of the polarized atoms with magnetic fields are observed \citep{Budker2007}. For instance, optical pumping of sodium with left-handed circularly polarized light produces atomic polarization in the $\ket*{F=2, m=+2}$ ground state (here $F$ is the total angular momentum in the ground state and $m$ is the Zeeman sublevel) that is depolarized by the action of magnetic fields due to spin-precession of the atomic angular momentum at the Larmor frequency \citep{Auzinsh2010}. If the medium is pumped with light pulses synchronized with the Larmor precession, a high degree of atomic polarization can be obtained and an increase in the fluorescence in the cycling transition $\ket*{ F=2, m=+2 } \rightarrow \ket*{F'=3, m'=+3}$ can be observed (the primed quantities refer to the excited atomic state) \citep{Bloom61}. The direct measurement of the Lamor frequency ($f_\text{Larmor}$) gives the magnetic field $B$ from $f_\text{Larmor}=\gamma B$, where $\gamma$ is the gyromagnetic ratio of ground-state sodium given by $\gamma=699\ 812$ Hz/G. This relationship applies to weak magnetic fields where Zeeman splitting of the energy levels depends linearly on the field. Therefore, as proposed in \citep{Higbie2011}, pumping the sodium layer with an intensity-modulated laser beam and observing the magneto-optical resonance occurring at the Lamor frequency from the surface of the Earth, allows one to remotely detect the magnetic field in the mesosphere. The first observation of a magnetic resonance and remote magnetic field determination in the mesosphere were recently reported in \citep{Kane2016}. Here, we demonstrate mesospheric magnetometry with an order-of-magnitude better sensitivity. In this work, the observed characteristic spectroscopic features of the resonance curve enable quantitative characterization of collisional processes in the mesosphere.

The fundamental sensitivity of an optical magnetometer is determined by the total number of atoms, the spin-relaxation rate in the atomic medium, and the measurement duration \citep{Budker2013}. At the Roque de los Muchachos Observatory (ORM), La Palma (Canary Islands), an average column density of $C_n=3.6\cdot 10^{13}$ atoms/m$^2$ has been measured with lidar observations \citep{Michaille2001}. From long-term observations of the sodium layer at low geographic latitude, the average sodium centroid height was determined to be 92 km above sea level with a thickness of 11.3 km \citep{Moussaoui2010}. The spin relaxation in the mesosphere is dominated by collisions and the finite transit time of the polarized atoms across the laser beam. Most collisions with sodium occur with N$_2$ and O$_2$ molecules, whereas Na--Na collisions are less frequent due to the low sodium density. While collisions of sodium with any molecule change the velocity of the atoms, Na--O$_2$ collisions are primarily responsible for spin relaxation due to the large exchange interaction between unpaired electrons of O$_2$ and sodium \citep{Morgan2010}. In absence of other spin-randomizing mechanisms, the rate of Na--O$_2$ collisions sets the fastest spin-relaxation rate in the atomic system and limits the sensitivity to magnetic field measurements by broadening the magneto-optical resonance. In the mesosphere, the transit time of sodium atoms across the laser beam is expected to be one order of magnitude longer than the relaxation time given by Na--O$_2$ collisions. 

The sensitivity can be affected by the instability of the sodium layer. The sodium atomic density in the mesosphere undergoes strong variability over all relevant time scales. Moreover, continuous monitoring of the sodium-layer density profiles with lidar (light detection and ranging) techniques shows structural and density changes with time scales of minutes. Particularly, sporadic events caused by the advection of meteor ablation from the ionosphere into the mesosphere produce sodium density changes in time scales of seconds \citep{Pfrommer2014,Clemensha2004}. In addition, atmospheric turbulence imposes another strong source of noise for an optical magnetometer. Even in an ideal case where there is no sodium density variation, the upper turbulent layers of the troposphere ($\sim$10 km altitude) heavily refract the wavefront of the downlink optical signal, creating a moving pattern of bright and dark areas on the surface. This effect is known as ``scintillation'' and it leads to intensity fluctuations of the observed light. The scintillation strength depends on the wavelength and angular extent of the source and on the aperture of the receiver telescope. The power spectrum of scintillation has been characterized at La Palma and shows a steep decrease for frequencies above 10 Hz for telescope apertures similar to those used in our experiment \citep{Dravins98}. 

In the following we describe the details for the experiment, present results, and discuss. In addition to magnetometry, our observations have yielded quantitative information about collisional processes in the mesosphere which is important for the optimization of sodium laser guide star and mesospheric magnetometers.


\section{Experiment}

\begin{figure}[t]
\centering
\includegraphics[width=0.4\textwidth]{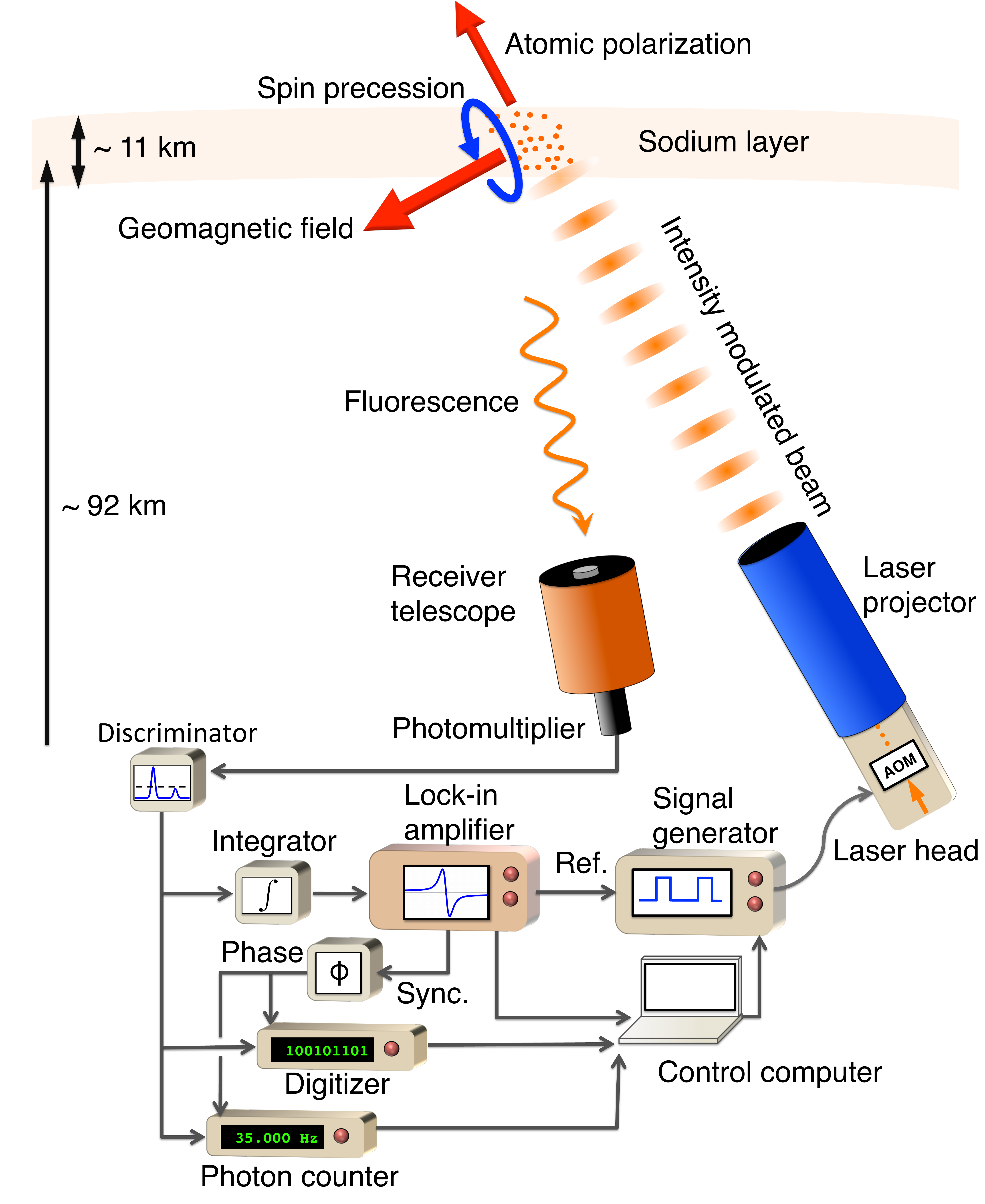}
\caption{Experimental arrangement for remote sensing of magnetic fields in the mesosphere. A laser projector sends an intensity-modulated beam to the mesosphere where it polarizes sodium atoms. Fluorescence is observed with a second telescope and the received photons are recorded, counted and demodulated with a digitizer, a photon counter, and a lock-in amplifier, respectively. The change in fluorescence is measured as the laser modulation frequency is swept around the Larmor frequency.}
\label{fig:Setup}
\end{figure}

The experimental setup is depicted in Figure \ref{fig:Setup}. It used the European Southern Observatory  Wendelstein Laser Guide Star Unit (ESO WLGSU) installed next to the William Herschel Telescope at ORM in La Palma. The operation of the WLGSU allows modulation of the beam and pointing the transmitter and receiver telescopes at the same target. The setup incorporated a laser projector telescope and a receiver telescope separated by eight meters. The light source consisted of a continuous-wave Raman-fiber-amplified frequency-doubled laser with a maximum output power of 20 W \citep{Bonaccini2012}. The laser was tuned to the vacuum wavelength of 589.158 nm corresponding to the $3S_{1/2} \rightarrow 3P_{3/2}$ transition of sodium (the D2 line); the linewidth of the laser was measured to be $\approx$2 MHz. The laser system incorporated an AOM (acousto-optic modulator) for amplitude modulation of the beam intensity. The beam polarization was controlled with a set of waveplates following the AOM. The Galilean projector telescope magnified the beam to an output diameter of 30 cm. The receiver consisted of a 40-cm aperture Schmidt-Cassegrain telescope mounted on the WLGSU receiver control unit, equipped with a narrow-band interference filter of 0.30(5) nm bandwidth centered at the sodium D2 line wavelength, a tracking CMOS (complementary metal–oxide–semiconductor) camera and a PMT (photomultiplier tube). A discriminator was used to filter and convert the analog pulses from the PMT into 100 ns TTL (transistor–transistor logic) pulses for the photon counters. The signal was detected by three independent data acquisition methods: a) digitizing and counting the arrival of individual photons (offline mode), b) directly measuring and averaging the photon-count difference per modulation period (online counter), and c) directly demodulating the signal from the PMT with a lock-in amplifier.

Each observation ``run" consisted of a discrete sweep of the laser modulation frequency around the predicted Larmor frequency. In order to reduce atmospheric scintillation noise, the rate of the laser pulsing ($f_\text{pulse}$) was continuously dithered with a square-wave function such that $f_\text{pulse}(t)=f_\text{step} + \delta f \cdot \text{sgn}\left[ \cos(2\pi f_m t) \right]$, where $f_\text{step}$ is the intensity modulation frequency at each step, $\delta f$ is the excursion amplitude, and $f_m$ is the dither frequency. The excursion was varied from $\delta f=8$ kHz to $\delta f =45$ kHz to find the optimal separation between demodulated peaks. The dither frequency was fixed at $f_m=150$ Hz for detection in a regime of low scintillation noise. Because of the probability that atoms decay into the dark $F=1$ ground state \citep{McClelland85}, a fraction of the laser power (12\%) was detuned by +1.712 GHz in order to maximize the number of available atoms by pumping them back into the $F=2$ ground state via the $F'=2$ excited state. The duty cycle of the laser intensity modulation was varied from 10\% to 30\%, as a compromise between high return flux and effective optical pumping. Laser polarization was kept circular for all runs in order to maximize the required atomic polarization along the laser beam direction \citep{Holz2010a}. The laser beam pointed in a direction at which the magnetic field vector in the mesosphere was approximately perpendicular to the laser-beam axis, which gives the highest contrast for the magneto-optical resonance. According to the World Magnetic Model (WMM2015) \citep{WMM2015}, the declination and inclination of the magnetic field at La Palma are 5.7\degree\ West and 39.1\degree\ downwards, respectively. Therefore, observations were carried out at an elevation of about 51\degree\ in the northern direction. Pointing at higher elevation up to 75\degree\ was also explored in order to reduce the airmass contribution to scintillation and the magnetic field uncertainty due to a shorter sodium layer path along the laser beam. From the WMM2015, the estimated magnetic field strength at 92 km altitude is 0.373 G, corresponding to a predicted Larmor frequency of 261 kHz. 

\begin{figure}[h]
\centering
\includegraphics[width=0.5\textwidth]{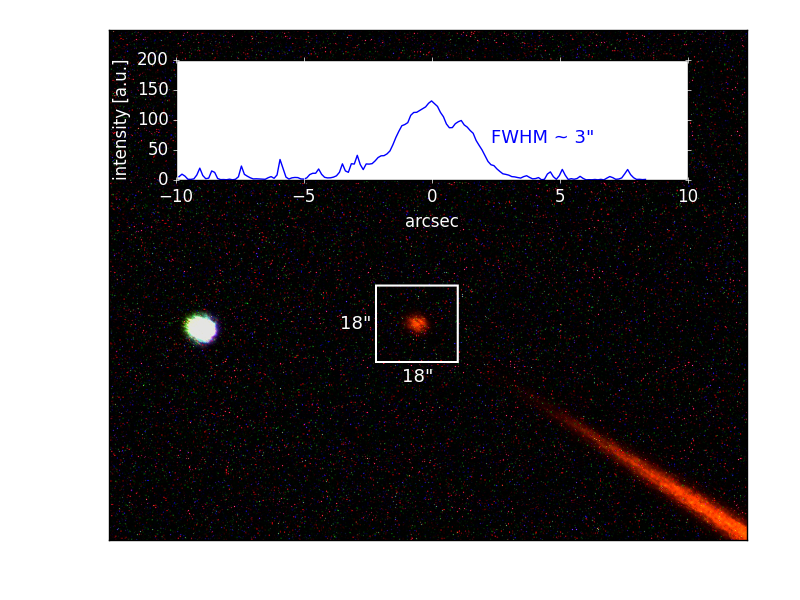}
\caption{A five-second-exposure image of the sodium fluorescence spot in the mesosphere along with the star HIP113889 obtained with the CMOS camera of the receiver telescope. The estimated long-term spot size is 3 arcsec, which comprises broadening due to atmospheric propagation and focusing error of the receiver telescope.}
\label{fig:LGS_intensity}
\end{figure}

The duration of each run depended on the frequency range of the sweep and the integration time for each step. About 10 minutes were necessary to perform a sweep of $\pm$75 kHz around the Larmor frequency. During five nights of observations, there were 51 successful runs. Laser power, duty cycle and excursion parameters were modified from run to run to investigate their effects on the magneto-optical resonance. The average atmospheric seeing was 0.7 arcsec at zenith as measured with a seeing Differential Image Motion Monitor (DIMM) collocated at the observatory. Data from the seeing monitor are available online from the website of the Isaac Newton Group of Telescopes (ING). Physical-optics modeling of the mesospheric spot size under these conditions \citep{Holz2010b} gives an instantaneous full-width-at-half-maximum (FWHM) beam diamater of $D_\text{FWHM}=36$ cm (0.8 arcsec) for a 30-cm launch telescope at an elevation angle of $\theta_\text{EL}=60$\degree, and average mesospheric irradiance of $I_\text{avg}^\text{meso}=15$ W/m$^2$ for 2-W CW output power (because of the duty cycle and finite AOM efficiency, 10--20 \% of the average laser power was deliverable to the sky). The spot size in the mesosphere was estimated from a long-exposure image taken with the receiver CMOS camera to be 3 arcsec (Fig. \ref{fig:LGS_intensity}), however, this estimation is subject to the effects of double-pass laser propagation through the atmosphere, beam-wander, and focusing error of the receiver, each of which contribute to make the apparent fluorescent spot broader than the instantaneous spot size. Since the spin-precession dynamics occurs on time scales of microseconds, we calculate the irradiance in the mesosphere using the instantaneous beam size obtained from physical-optics modeling.

\section{Results}

\begin{figure*}[h]
\centering
\includegraphics[width=1\textwidth]{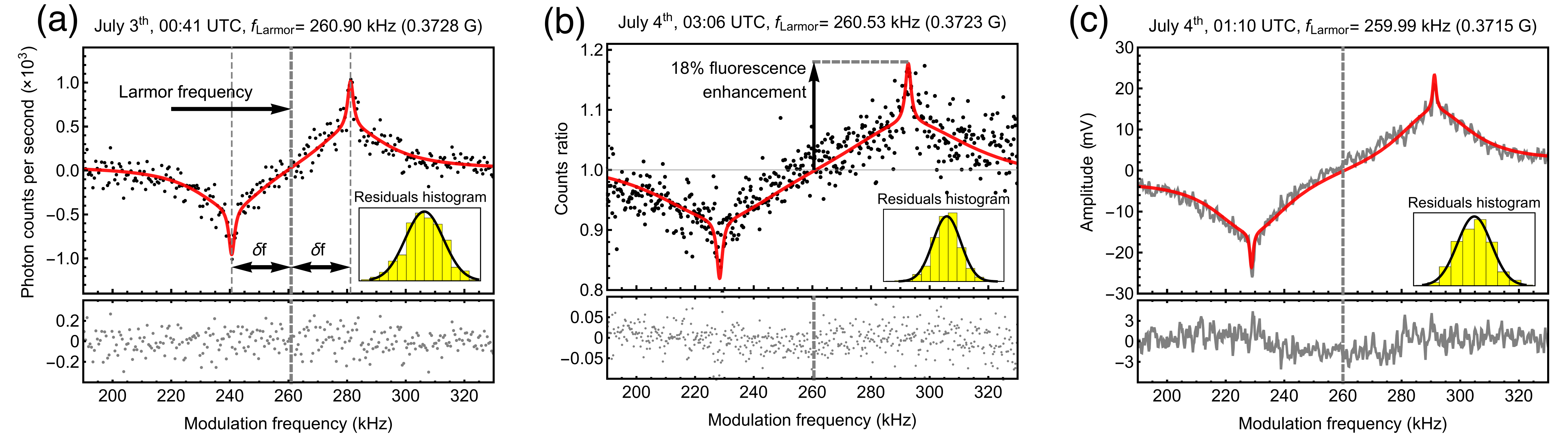}
\caption{Magneto-optical resonances obtained by sweeping the frequency of the intensity modulated laser beam with three concurrent data acquisition methods. (a) Online differential counter for a modulation duty cycle of 20\% and $I_\text{avg}^\text{meso}=13$ W/m$^2$. The Larmor frequency lies in the center between the peaks, which are separated by twice the dither excursion $\delta f=20.2$ kHz. (b) Ratio of the photon counts per dither period averaged over 2 s. The modulation duty cycle was 30\%, excursion $\delta f=30.8$ kHz and calculated mesospheric irradiance $I_\text{avg}^\text{meso}=33$ W/m$^2$. (c) Lock-in amplifier with time constant of 300 ms, modulation duty cycle 20\%, excursion $\delta f=30.8$ kHz, and calculated mesospheric irradiance $I_\text{avg}^\text{meso}=17$ W/m$^2$. For all resonances a double Lorentzian fit shows a broad and a narrow width of approximately 30 kHz and 2 kHz respectively, consistent with two relaxation mechanisms due to velocity-changing collisions and spin-exchange collisions of sodium with N$_2$ and O$_2$ molecules. The residuals of the fits are shown below each resonance and obey a normal distribution.}
\label{fig:Resonances}
\end{figure*}

Figure \ref{fig:Resonances} shows three typical demodulated signals obtained with an online differential counter (Fig. \ref{fig:Resonances}a), an offline ratio counter (Fig. \ref{fig:Resonances}b), and a lock-in amplifier (Fig. \ref{fig:Resonances}c). The online counter reported the real-time difference in the photon counts between two half-periods of the dither signal, averaged over the time of each frequency step (2--3 s). At a dither frequency of 150 Hz, the averaged difference between off-resonance and on-resonance is about 1000 counts/s. For each dither period (6.7 ms), the count difference is only about six photon counts. A higher dither frequency would have rejected scintillation noise better, at the cost of fewer photon counts per dither period. The digitizer recorded all photon counts and the ratio between alternating dither sub-periods was calculated. With this estimator, drifts in the photon flux due to sodium abundance variations during the observation were suppressed. During post-processing, the phase of the square-wave dither signal could be freely adjusted. This is in contrast to the case of the online counter, where a wrong input phase could suppress the signal without the possibility of recovering it in post-processing. In addition, the lock-in amplifier demodulated the incoming signal into phase and quadrature components, calculating in real time the time-evolution of the resonance. A time constant of 300 ms was used for all measurements. 

The signal consisted of a positive and a negative peak, each comprising two superimposed Lorentzians (Fig. \ref{fig:Resonances}). The Larmor frequency lies at the mid-point between the two peaks. The fit residuals are shown below each resonance as well as their histogram of residuals. The residuals from the lock-in amplifier signal display small deviations from the fit that may be attributable to slow altitude displacements of the sodium layer centroid during the performance of the sweep. Upward displacement of the sodium centroid toward a weaker magnetic field region produce a shift of the magnetic resonance toward lower frequencies, resulting in asymmetries of the observed resonance.

\begin{figure}[h]
\centering
\includegraphics[width=0.5\textwidth]{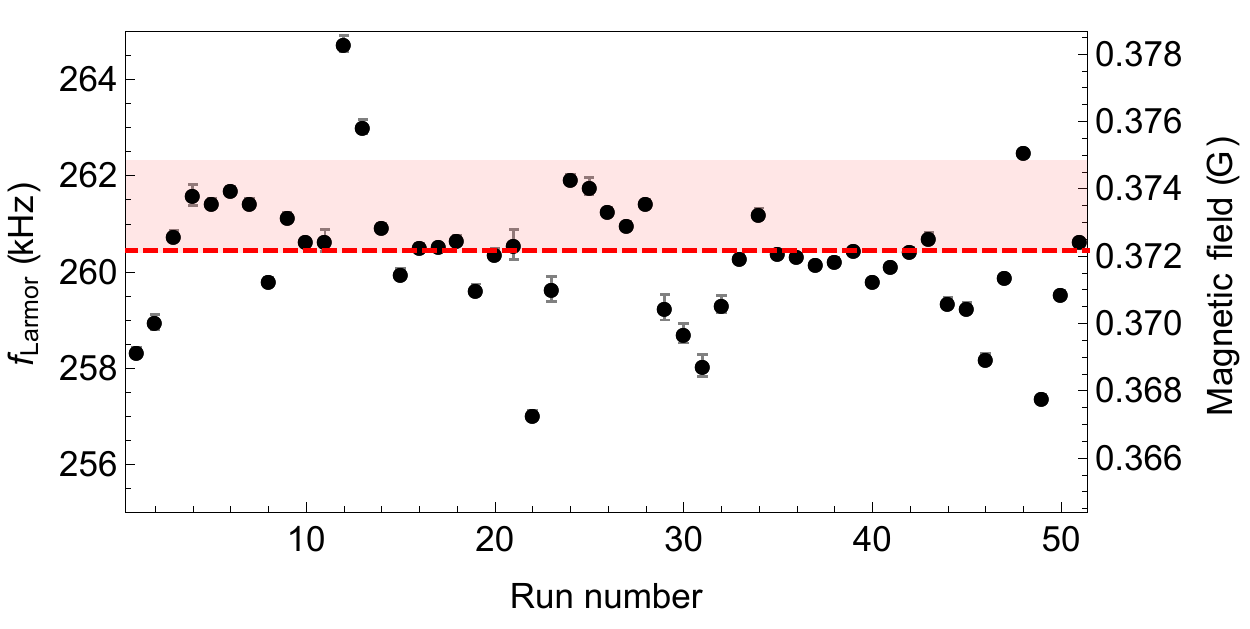}
\caption{The measured Larmor frequency (geomagnetic field) from 51 runs. The red dashed line is the mean of all observations. The horizontal light-red band represents the predicted magnetic field between 85 km and 100 km altitude according to the WMM2015 magnetic model. Error bars are the standard error of the estimate of $f_\text{Larmor}$.}
\label{fig:Larmor_frequency}
\end{figure}

The results for all 51 runs are plotted in Fig. \ref{fig:Larmor_frequency}. The average Larmor frequency was found to be 260.4(1) kHz, representing a geomagnetic field of 0.3720(1) G. The WMM2015 prediction for the magnetic field at 92 km altitude is 0.3735(15) G, giving a difference of less than 0.5\% between the model and our observations. With this technique, the accuracy of the determination of the magnetic field depends on the photon shot noise, atmospheric scintillation, as well as random fluctuations of the centroid and the sodium layer profile. The spatial resolution is constrained by the elevation at which the observation is carried out. Low elevation measurements lead to an extended probing volume in the mesosphere and therefore a larger sample area in the horizontal plane. For this reason, the spatial resolution for mapping magnetic fields is limited to a horizontal trace of about 6.5 km for an elevation of $\theta_\text{EL}=60$\degree, given a median sodium layer thickness of 11.3 km at zenith. The main geomagnetic field gradient in the vertical direction $H$ in the mesosphere is $dB/dH=-1.85\times 10^{-4}$ G/km \citep{WMM2015}, equivalent to a Larmor frequency gradient of $df_\text{Larmor}/dH=-0.129$ kHz/km. This field gradient shifts the center of the resonance peak and broadens it, imposing an accuracy limitation on this method. 

\begin{figure*}[h]
\centering
\includegraphics[width=0.45\textwidth]{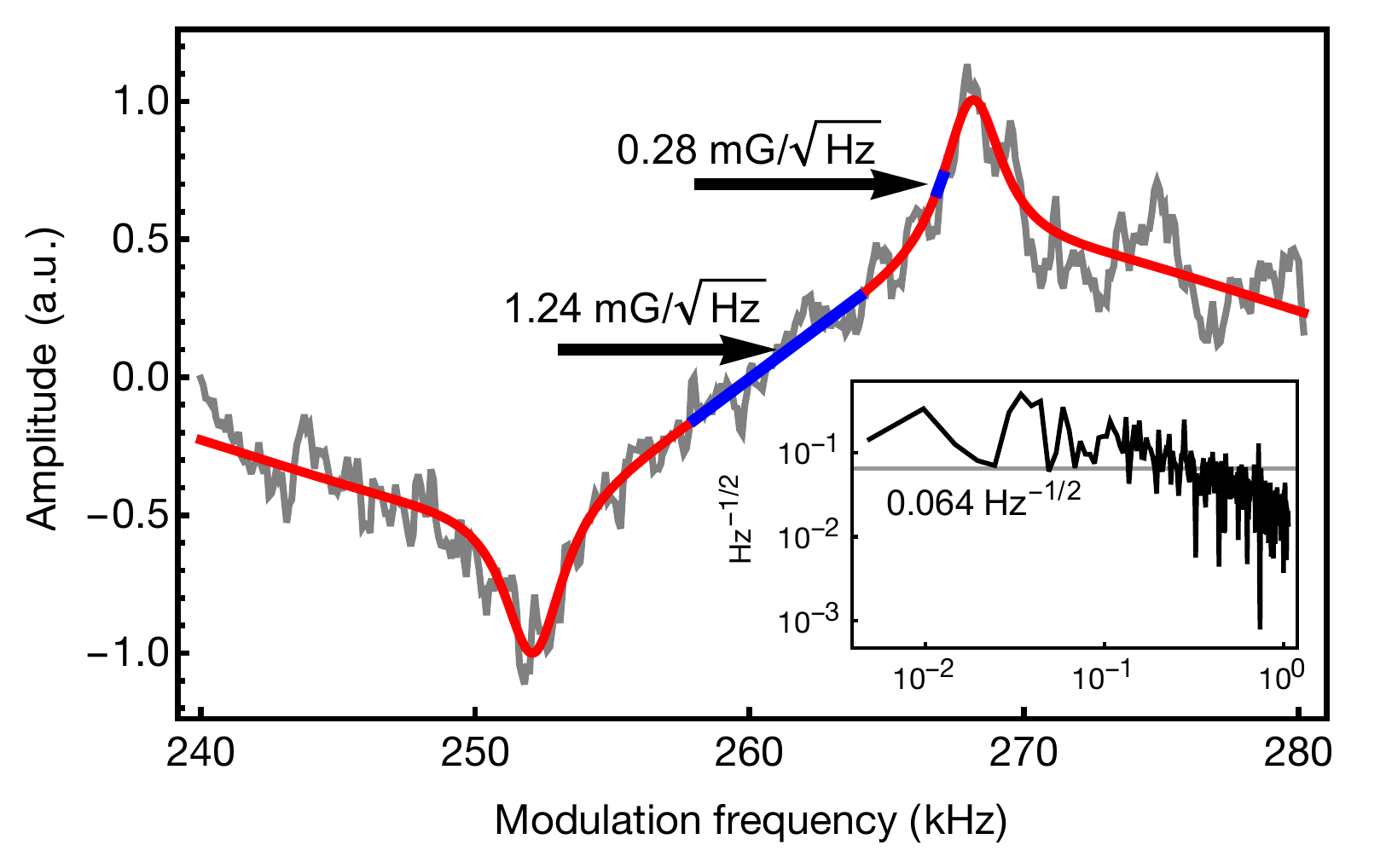} 
\includegraphics[width=0.45\textwidth]{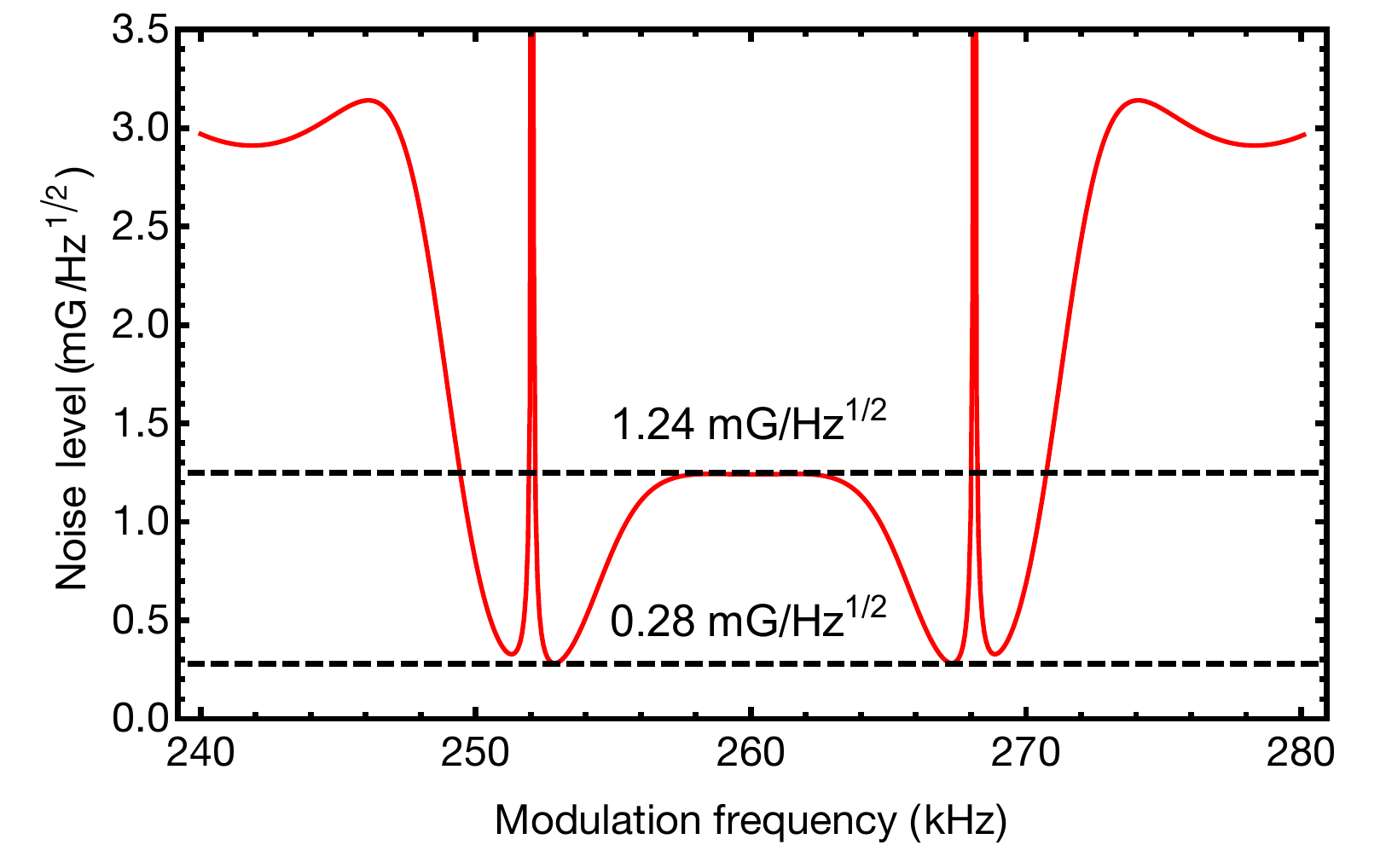}
\caption{Estimation of the magnetometry accuracy along the magneto-optical resonance. The average amplitude of the spectral density of residuals is 0.064 Hz$^{-1/2}$. The maximum accuracy is achieved in the steepest points of the narrow component resonance slope.}
\label{fig:Noise_floor}
\end{figure*}

Spatially separated sodium density peaks produced by sporadic events and/or sporadic sodium layers \citep{Clemensha94} broaden the magneto-optical resonance. Sporadic events in La Palma have been detected in average one per night with lifetime from 30 seconds to several hours \citep{Michaille2001}. The accuracy could be improved if the vertical sodium profile were independently known, for example, from simultaneous lidar measurements. The magnetometry accuracy is estimated for a good magneto-optical resonance profile obtained with excursion $\delta f=8$ kHz (Fig. \ref{fig:Noise_floor}). 

To measure the absolute magnetic field in the mesosphere, a full scan of the magneto-optical resonance was performed so that the Larmor frequency could be determined. On the other side, to measure fluctuations in the magnetic field, the magnetometer can operate with a modulation frequency fixed at the maximum sensitivity point along the resonance curve. The maximum sensitivity point can be found at the maximum of the differentiated fit function of the resonance. The density spectrum of residuals is used to quantify all noise contributions in the system and to calculate the magnetometry accuracy at each point of the resonance.

At the middle point between the two peaks the calculated accuracy is 1.24 mG/$\sqrt{\text{Hz}}$, similar to that reported in \citep{Kane2016}. The presence of the narrow resonance peaks with larger magnetic response than the broad component (steeper slope) provides higher-sensitivity points where an accuracy of 0.28 mG/$\sqrt{\text{Hz}}$ can be reached. Figure 5 shows the calculated magnetometry accuracy along the magneto-optical resonance. The standard error of estimating $f_\text{Larmor}$ for the resonance in Fig. \ref{fig:Noise_floor} is 0.04 kHz (0.05 mG). The existence of the narrow peaks found in this experiment decreases the uncertainty in the estimate of the Larmor frequency.

\section{Discussion}

\begin{figure}[h]
\centering
\includegraphics[width=0.4\textwidth]{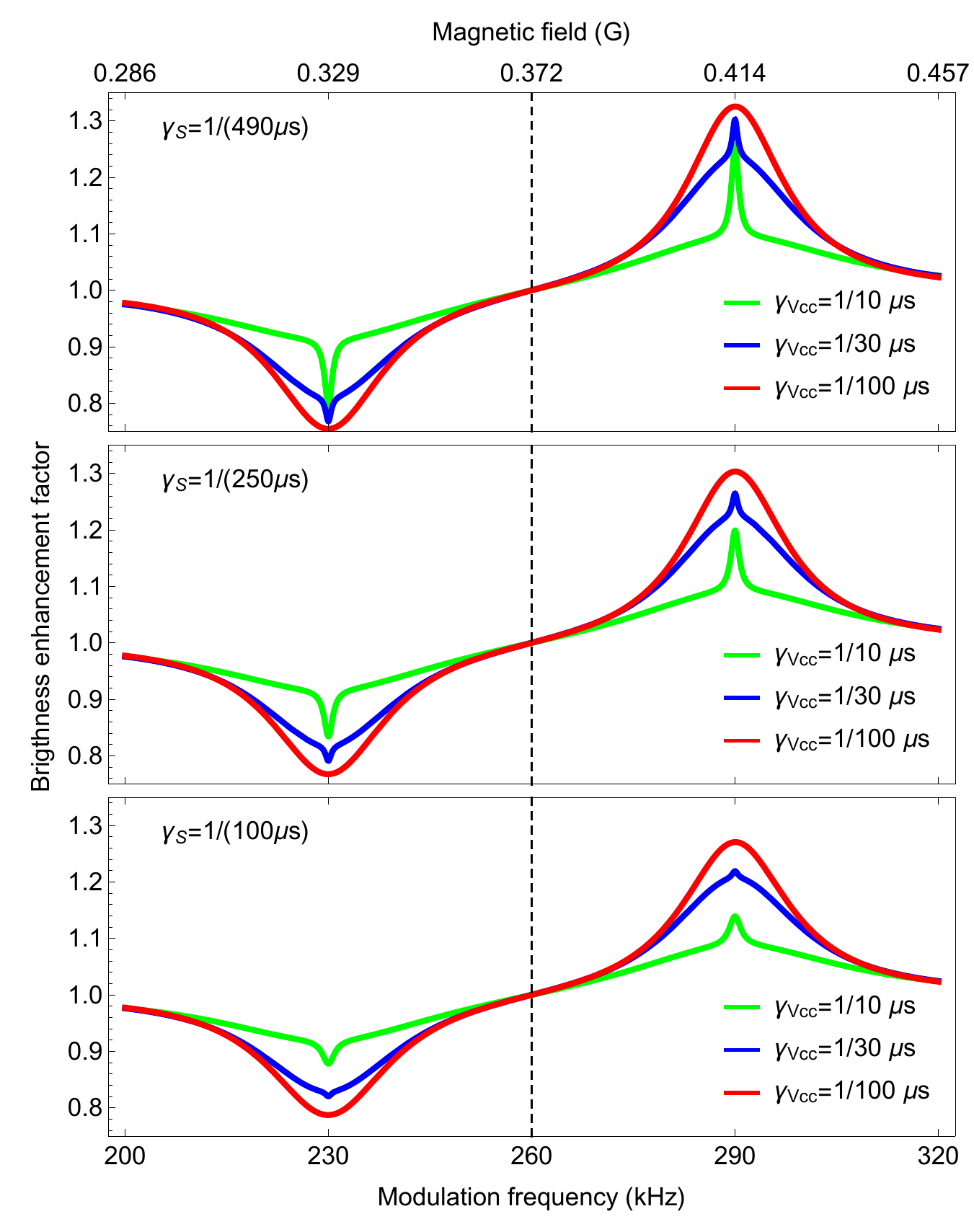}
\caption{Numerical modeling of the magneto-optical resonance for different collision relaxation rates, assuming a laser intensity of $I_\text{avg}^\text{meso}=15$ W/m$^2$ in the mesosphere, and excursion of $\delta f =30$ kHz. The central dashed line indicates the Larmor frequency and the equivalent magnetic field.}
\label{fig:Model}
\end{figure}

Numerical modeling of the time evolution of the sodium atomic polarization under resonant pulsed excitation using the density-matrix model described in \citep{Rochester2012} shows a magnetic resonance that can be fit with two superimposed Lorentzians of different widths. The superimposed Lorentzians (see also \citep{Fan2016}) are analogous to the nested dispersive Lorentzians observed in nonlinear magneto-optical rotation (NMOR) with antirelaxation-coated vapor cells \citep{Budker2002}. In such NMOR experiments, a ``transit" effect is observed with a resonance width corresponding to the rate at which atoms traverse the light beam, while a narrower ``wall" effect --- due to atoms leaving and then reentering the light beam after bouncing off of the cell wall --- has a width corresponding to the relaxation rate of atomic ground-state polarization. In the present case, rather than considering the atomic positions relative to the light beam, we must consider the atomic velocities as they leave and reenter the resonant velocity group of the Doppler distribution; the transit effect then corresponds to a velocity-changing collision effect, while the wall effect corresponds to an effect with a narrower resonance width determined by the polarization relaxation rate due to spin-exchange collisions. The effect of varying the collision rates is seen in simulated resonance curves obtained from our model (Fig. \ref{fig:Model}). To a first approximation, we have assumed that the diffusion process between velocity groups is independent of the distance in velocity classes that atoms jump, even though the real scenario is  complex and requires more sophisticated modelling of the effect of velocity-changing collisions \citep{McGuyer2012}.


Fitting the experimental data with a double Lorentzian function and estimating the widths yields FWHM median values of $\Delta f_\text{broad}=32$ kHz for the broad resonance component, and $\Delta f_\text{narrow}=2.4$ kHz for the narrow resonance component. According to our numerical simulations the observed widths are obtained with velocity-changing collision rate on the order of $\gamma_\text{vcc} \approx 1/(10\ \mu \text{s})$ and a spin-exchange rate on the order of $\gamma_\text{s} \approx 1/(100\ \mu \text{s})$. These results suggest that collision rates in the mesosphere are higher than previous estimates by a factor of 2--6. For instance, a mean spin-exchange collision rate of  $\gamma_\text{s} = 1/(490\ \mu \text{s})$ was estimated in \citep{Holz2010a}. Other estimates suggest values of $\gamma_\text{s}  \approx 1/(200\ \mu \text{s})$ \citep{Lihang2016}, and $\gamma_\text{s} = 1/(640\ \mu \text{s})$\citep{Miloni99}. While other methods to estimate the sodium spin-exchange collision rate depend on estimates of the atomic cross-section between Na and other species, our estimate provides a relatively direct measurement of $\gamma_\text{s}$. This value could be used, based on first principles, to calculate the actual Na--O$_2$ cross-section in the mesosphere (to the best of our knowledge no experimental measurement of the Na--O$_2$ spin-exchange cross-section at mesospheric conditions has been reported).

The discrepancy between collision rates estimates may be due to bias in the assumed cross-sections, large magnetic field gradients, and/or uncertainty in the sodium profile. In order to identify the reason for this discrepancy, quantitative measurements, development of an improved collision model, and parallel sodium profile measurements with lidar could be used.

\section{Conclusion}

We have demonstrated a method of remote magnetic field measurements in the mesosphere using a laser beam with intensity modulation at the Larmor frequency of sodium, achieving an accuracy of 0.28 mG/$\sqrt{\text{Hz}}$. This work contributes to several efforts in the scientific community to develop techniques for remote sensing of magnetic fields in the atmosphere \citep{Kane2016,Luke2014,Magnar2016,Remas89}. We note that the setup used in this experiment can, in principle, be realized with components such as laser sources, modulators, and telescopes currently available commercially. Our observations show good agreement with the predictions of the geomagnetic field from the World Magnetic Model for altitudes between 85 km and 100 km, and could provide input data for future assessments of this model. We found that the magneto-optical resonant signal contains broad and narrow features that depend on specific kinds of atomic collisions. The method presented in this work shows that atomic collision rates can be inferred from the observed resonances, suggesting another important application of this approach: remote sensing of collisional processes in the mesosphere.\\

\noindent \textbf{Data availability.} The data that support the findings of this study may be available to qualified scientists from the corresponding author upon request. 

\pagebreak

\subsection*{Acknowledgements}

This work was partly supported by the Office of Naval Research Global under grant N62909-16-1-2113. F.P.B. acknowledges the support from a Carl-Zeiss Foundation Doctoral Scholarship. P.H. gratefully acknowledges financial support from the Natural Sciences and Engineering Research Council of Canada. We thank the Instituto de Astrofísica de Canarias (IAC) for their support during the measurements campaign at the Observatorio Roque de los Muchachos. We thank the workshops of the Department for Physics and Astronomy of UBC for their contribution to the hardware development. 

\subsection*{Author contributions}

F.P.B. designed and carried out the experiment, processed and analyzed data, wrote the manuscript.
D.B.C. participated in the experimental setup definition, modified the ESO WLGSU for the experiment, operated the WLGSU during the experiment. 
D.B. developed the experimental concept and procedure and interpreted the results.
M.C. modified the ESO WLGSU for the experiment, operated the WLGSU during the experiment. 
J.H. developed parts of the receiver, carried out the experiment, processed data.
P.H. designed and built the receiver.
R.H. performed data analysis, numerical simulation and error estimation.
S.R. developed the computational model and participated in the theoretical analysis.
All co-authors contributed to and reviewed the manuscript.

\subsection*{Material and correspondence}

Address any request of information to Felipe Pedreros Bustos by email at pedreros@uni-mainz.de

\subsection*{Additional information}

\textbf{Competing interests:} The authors declare no competing financial interest.

\section*{Additional material}

\begin{figure*}[h]
\centering
\includegraphics[width=0.48\textwidth]{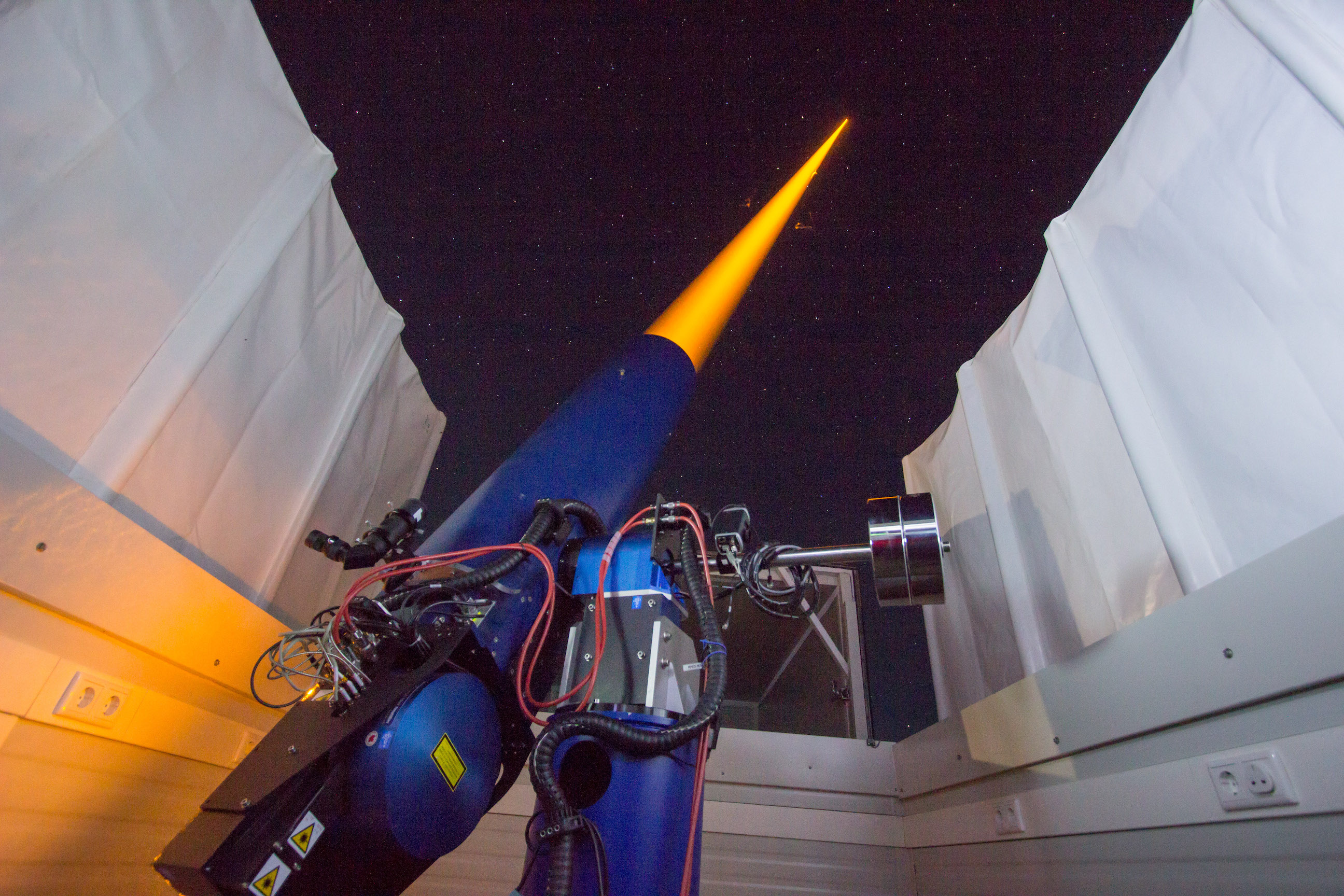} 
\includegraphics[width=0.48\textwidth]{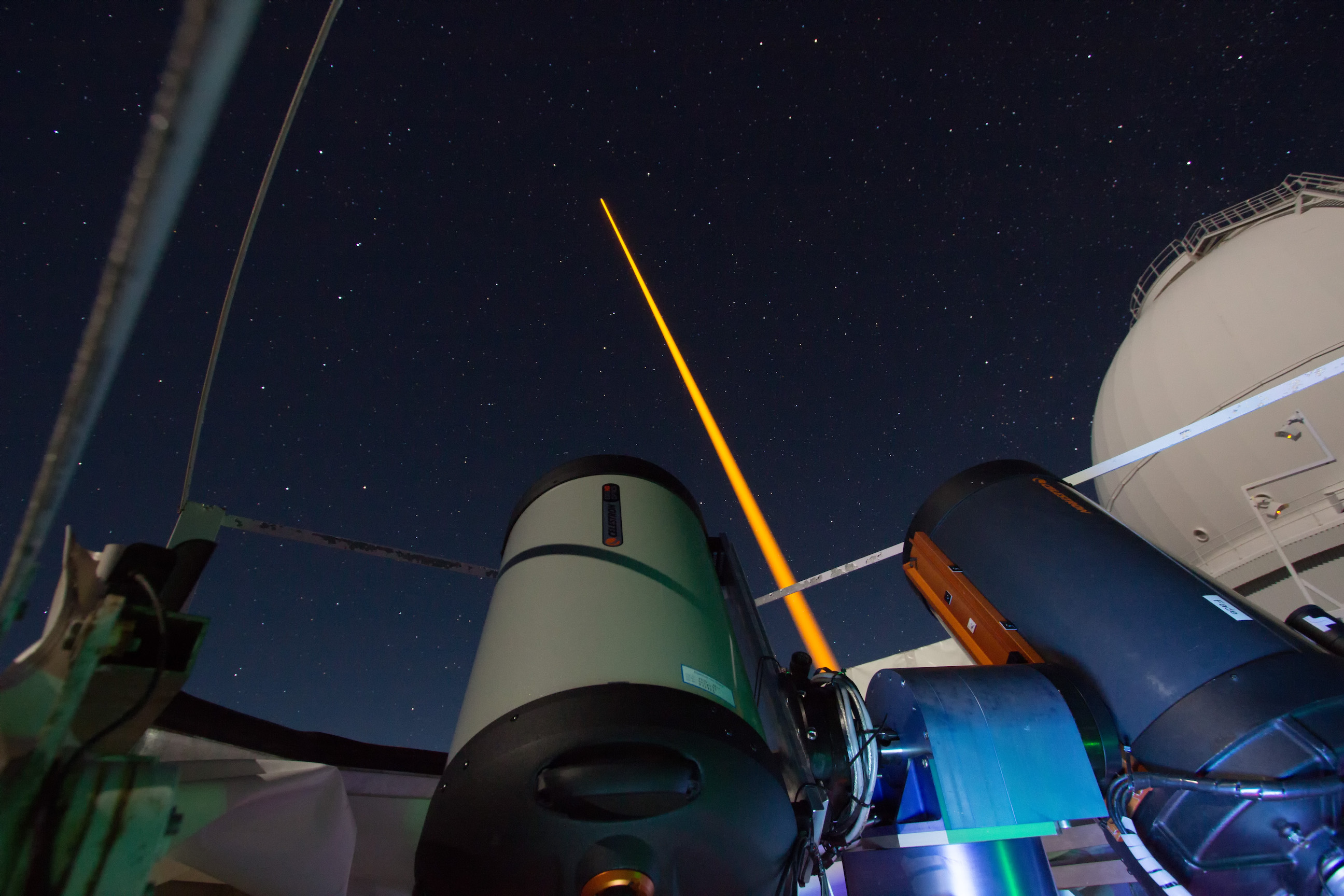}
\caption{(Left) The 20-Watt laser projector of the European Southern Observatory laser guide star system. (Right) The observing telescope placed next to the laser projector for detection of LGS brightness enhancement.}
\label{fig:WLGSU}
\end{figure*}

\end{document}